\documentclass[numberedappendix,apjl]{emulateapj}

\usepackage{apjfonts}
\usepackage{color}

\begin{document}

\title[]{The effect of unresolved binaries on globular cluster proper-motion dispersion profiles}

 \author{P. Bianchini\altaffilmark{1,$\star$} \affil{Max-Planck Institute for Astronomy, K\"{o}nigstuhl 17, 69117 Heidelberg, Germany; bianchini@mpia.de}, M. A. Norris\altaffilmark{1,3}, G. van de Ven\altaffilmark{1}, E. Schinnerer\altaffilmark{1}, A. Bellini\altaffilmark{2}, R. P. van der Marel\altaffilmark{2}, L. L. Watkins\altaffilmark{2}, J. Anderson\altaffilmark{2} } 
 
\altaffiltext{1}{Max-Planck Institute for Astronomy, K\"{o}nigstuhl 17, 69117 Heidelberg, Germany; bianchini@mpia.de} 
\altaffiltext{2}{Space Telescope Science Institute, 3700 San Martin Drive, Baltimore, MD 21218, USA} 
\altaffiltext{3}{University of Central Lancashire, Preston, PR1 2HE, UK}
 \altaffiltext{$\star$}{Member of the International Max Planck Research School for Astronomy and Cosmic Physics at the University of Heidelberg, IMPRS-HD, Germany.}

\begin{abstract}
High-precision kinematic studies of globular clusters require an accurate knowledge of all possible sources of contamination. Amongst other sources, binary stars can introduce systematic biases in the kinematics. Using a set of Monte Carlo cluster simulations with different concentrations and binary fractions, we investigate the effect of unresolved binaries on proper-motion dispersion profiles, treating the simulations like \textit{HST} proper-motion samples. Since globular clusters evolve towards a state of partial energy equipartition, more massive stars lose energy and decrease their velocity dispersion. As a consequence, on average, binaries have a lower velocity dispersion, since they are more massive kinematic tracers.
We show that, in the case of clusters with high binary fraction (initial binary fraction of 50\%) and high concentration (i.e., closer to energy equipartition), unresolved binaries introduce a color-dependent bias in the velocity dispersion of main-sequence stars of the order of 0.1-0.3 km s$^{-1}$ (corresponding to $1-6$\% of the velocity dispersion), with the reddest stars having a lower velocity dispersion, due to the higher fraction of contaminating binaries. This bias depends on the ability to distinguish binaries from single stars, on the details of the color-magnitude diagram and the photometric errors. We apply our analysis to the HSTPROMO data set of NGC 7078 (M15) and show that no effect ascribable to binaries is observed, consistent with the low binary fraction of the cluster. Our work indicates that binaries do not significantly bias proper-motion velocity-dispersion profiles, but should be taken into account in the error budget of kinematic analyses.

\end{abstract}
 

\section{Introduction}

\begin{table*}
\begin{center}
\caption{}
\begin{tabular}{llccccccccc}
\hline\hline
Simulation name &initial concentration &$f_\mathrm{initial}$ & $f_\mathrm{final}$ & $f_\mathrm{FoV}$ & $f_\mathrm{R_c}$ & \multicolumn{2}{c}{R$_c$} & \multicolumn{2}{c}{R$_h$} & c\\
&&&&&&pc&arcmin&pc&arcmin&\\
\hline
Simulation 1&intermediate&10\%&2.9\%& 5.5\%& 7.3\%&3.15 &2.16&4.92 & 3.38 &1.45\\
Simulation  2&intermediate&50\%&16.2\% &28.5\% &30.8\%&3.89 & 2.68&6.06 & 4.16 &1.34\\

Simulation  3&low &10\%&5.5\% &9.6\% & 9.5\% &6.07 & 4.17&9.05 & 6.22 & 1.16\\
Simulation  4&low&50\%&20.2\% &32.6\% & 31.5\%&6.47 & 4.45&10.92 & 7.51& 1.12 \\

Simulation  5&high&10\%&3.4\% & 6.8\%&16.4\%&0.75 & 0.51 &2.69 & 1.85 & 2.06\\
Simulation  6&high&50\%&11.5\% & 21.1\%& 32.0\%& 1.34 & 0.92 & 3.05 & 2.09& 1.79 \\

\hline
\end{tabular}
\tablecomments{Properties ofthe simulations for a 11 Gyr snapshot observed at 5 kpc. The name of the simulation is followed by the primordial binary fraction $f_\mathrm{initial}$, the global binary fraction at 11 Gyr $f_\mathrm{final}$, the binary fraction within the FoV of 4 arcmin and magnitude cut between 17 and 24 $r$-band mag, and the binary fraction within the core radius $R_c$. The concentration $c=\log (R_t/R_c)$, the core radius $R_c$, and the half-light radius $R_h$ are also provided.}
\label{tab:1}
\end{center}
\end{table*}

Globular clusters (GCs) are some of the oldest stellar systems in the universe and provide crucial information on the early phases of galaxy formation and assembly. In order to constrain their formation, a growing focus has been devoted to the study of their internal kinematics, which provides a long-lasting fossil record of their formation and dynamical evolution.

Typical kinematic studies of Galactic GCs are based on line-of-sight velocities, from spectroscopic measurements either of resolved stars, or of (partially) unresolved stars through integrated-light spectroscopy. These measurements are limited by the fact that only one component of the velocity-vector is observed and only for the brightest sources, i.e., giant stars all with similar mass. However, a significant improvement was recently possible thanks to high-precision \textit{Hubble Space Telescope} (\textit{HST}) proper motions (HSTPROMO, \citealp{Bellini2014, Watkins2015,Watkins2015b}), providing velocity measurements on the plane-of-the-sky for 22 GCs, with a median of $\sim$60,000 stars per cluster, and thanks to other proper-motion samples dedicated to single GCs (e.g., \citealp{McLaughlin2006,Richer2013} for NGC 104, \citealp{McNamara2003} for NGC 7078, \citealp{McNamaraMcKeever2011,McNamara2012} for NGC 6266, and \citealp{Anderson2010} for NGC 5139).

Proper-motion samples provide two-dimensional velocity information (the two components on the plane-of-the-sky) and sample both bright giant and fainter stars along the main sequence, allowing to measure the motion of stars with different masses. Moreover, since they provide large samples of stars, they are ideal to study the detailed internal kinematics of a GC, reaching low levels of random errors, and possibly allowing the coupling of the kinematics with color and chemical information. 

An accurate kinematic analysis requires that any bias present in the observed kinematics should be well understood. A common source of contamination is binaries. Binaries can contaminate line-of-sight measurements, due to the motions of the stellar components around their mutual barycenter adding to the systemic motion of the binary system (e.g., \citealp{Minor2010}). Unidentified binaries can cause an overestimate of the measured line-of-sight velocity dispersion. Unfortunately, their identification is challenging, even for bright stars, since it requires repeated spectroscopic measurements. The resulting bias is negligible in typical GCs, characterized by a low binary fraction (lower than that of field stars, \citealp{Milone2012}), but it can be crucial for faint stellar systems (e.g., ultra-faint dwarf galaxies) with low velocity dispersion and small data samples \citep{McConnachie2010, Bradford2011}. 

In the case of proper motions, binaries can affect the data in three ways. First, the internal binary motion can produce an effect on the astrometric measurements, adding scatter to the proper-motion determination; this is, however, negligible.\footnote{The effect is proportional to $a/(\sigma\,T)$, with $a$ major axis of the binary, $\sigma$ typical velocity dispersion, $T$ time-baseline of the proper-motion measurements; this effect is maximum for face-on orbits and epochs separated by a half-integer of the orbital period.} Second, semi-resolved binaries seen as astrometric blends, can be characterized by poor-quality astrometry (because of the difficulties in centroiding the PSF), that translates to biased proper-motion measurements. This effect can be reliably mitigated using quality selection criteria on the proper-motion samples (see the discussion for NGC 7078; \citealp{Bellini2014}). Third, unresolved binaries can bias the kinematics since they are a more-massive kinematic tracer than single stars at a given magnitude. In fact, due to two-body interactions, GCs evolve towards a state of partial energy equipartition \citep{TrentivanderMarel2013,Bianchini2015d}. This means that while more massive stars lose energy and sink towards the center, less massive stars gain energy and move outwards, leading to a mass-dependent kinematics.

Unresolved binaries, being more massive kinematic tracers, will be characterized by a lower velocity dispersion.\footnote{Such mass-dependent kinematics can also be observed by comparing blue straggler stars against evolved stars (\citealp{Baldwin}).} The effect of binaries on proper-motions samples would therefore be to systematically decrease the velocity dispersion otherwise measured from single stars alone. Moreover, since unresolved binaries are seen as a single star with flux given by the sum of the fluxes of the two components, they are redder in color along the main sequence. Their kinematic effect could therefore introduce a subtle color-dependence on the kinematics, that must be disentangled from any other possible color-dependent kinematic effects (e.g., the kinematic differences expected from different multiple stellar population scenarios, \citealp{Henault-Brunet2015} and the observations presented in \citealp{Richer2013} for NGC104 and \citealp{Bellini2015} for NGC2808).

The aim of this letter is to understand and quantify the kinematic effect of unresolved binaries connected to partial energy equipartition. We use a two step approach: firstly, we analyze a set of Monte Carlo cluster simulations \citep{Downing2010} spanning a realistic range of concentrations and binary fractions, treating them as mock \textit{HST} proper-motion samples to look for the kinematic effects of binaries (Section \ref{sec:2} and \ref{sec:3}). Secondly, we apply the same analysis to the HSTPROMO data of NGC 7078 (M15) for a comparison to observations (Section \ref{sec:4}).

\begin{figure*}
\centering
\includegraphics[width=1\textwidth]{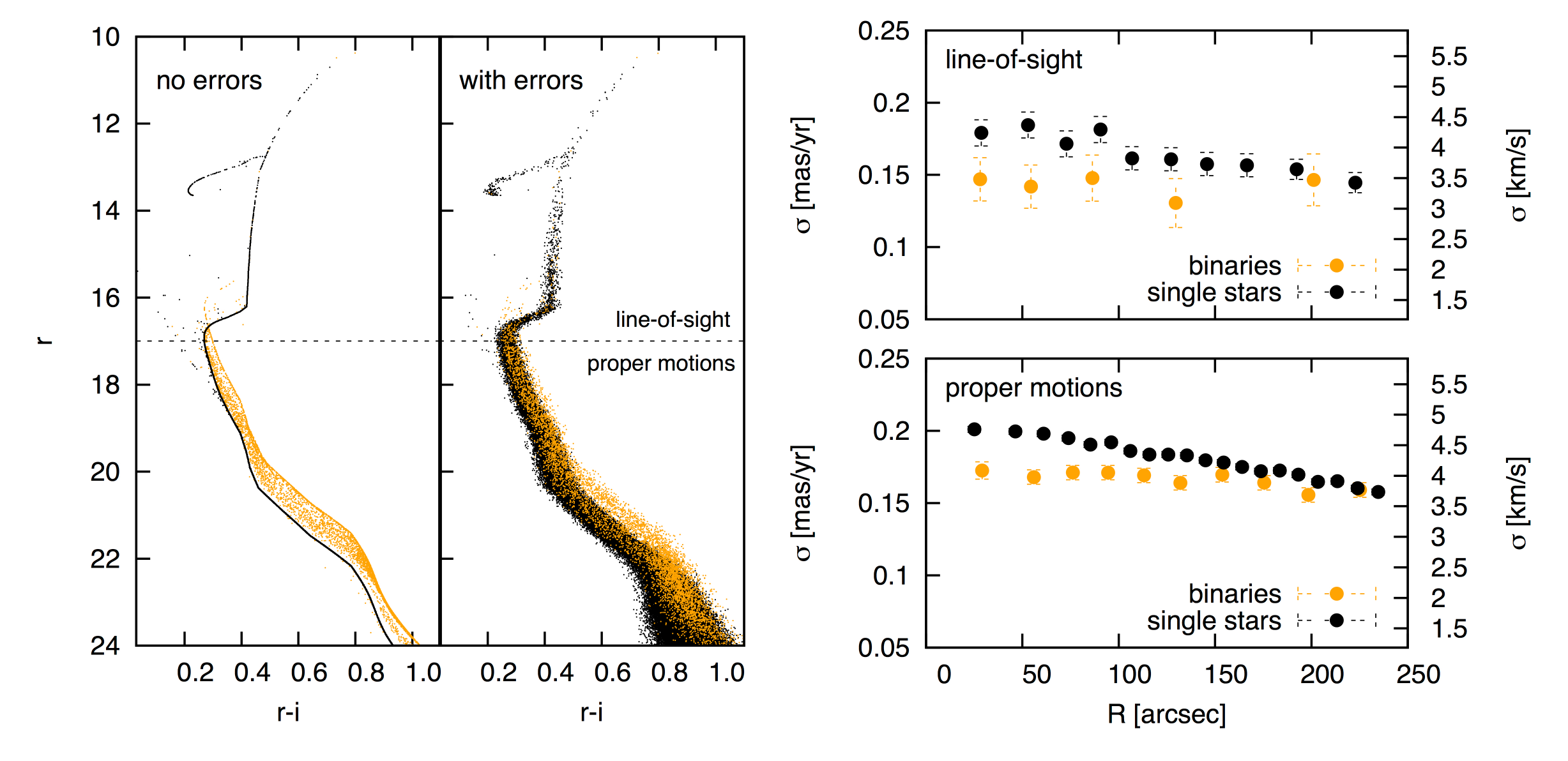}
\caption{\textbf{Left:} Color magnitude diagrams of Simulation 1, with and without artificial errors. Binaries (orange) are in a redder sequence parallel to the main sequence. Note that the simulated r vs. r-i diagram does not match closely the typical observed r v. r-i diagram. \textbf{Right:} Velocity-dispersion profiles for single and binaries separately. The top panel refers to bright stars only, for which typical measurements of line-of-sight velocities are available, while the bottom panel refers to a typical magnitude range sampled by proper motions. Binaries show a lower velocity-dispersion profile (globally, 8\% lower than single stars) because of the effect of partial energy equipartition.}
\label{fig:2}
\end{figure*}

\section{Simulations}
\label{sec:2}

We consider a set of Monte Carlo cluster simulations, developed by \citet{Downing2010} with the Monte Carlo code by \citet{Giersz1998}. The simulations have an initial number of particles of $N=500\,000$, an initial mass function (IMF), stellar evolution, and primordial binaries, providing a realistic description of the long-term evolution of GCs with a single stellar population. 

All simulations have initial conditions drawn from a \citet{Plummer1911} model, a \citet{Kroupa2001} IMF, [Fe/H]=$-$1.3, and an initial tidal cut-off at 150 pc (making them relatively isolated). We consider 6 simulations, characterized by 3 values of initial concentration with binary fraction of 10\% or 50\%. The initial binary parameters are described in \citet{Downing2010} and are based on the eigenvalue evolution and feeding algorithms of \citet{Kroupa1995}. These prescriptions use birth mass ratios drawn at random from the \citet{Kroupa1993} IMF. After 11 Gyr of dynamical evolution, the distribution of the mass-ratio $q$ is flat for $q>0.5$ (consistent with observations, e.g., \citealp{Milone2016}), decreasing for $q<0.5$, and with an excess of binaries at $q\approx1$. The output of each simulation consists of the three-dimensional position and velocity vectors, the stellar mass and the magnitude for each star (in $r$ and $i$ Sloan bands).  

We treat our simulations similarly to real observations (in particular to HSTPROMO proper motions, \citealp{Bellini2014}), considering the properties of an 11 Gyr snapshot, projected along the line-of-sight at 5 kpc distance (typical of a Galactic GC). Furthermore, we consider stars in a field-of-view (FoV) of 4 arcmin around the center of the cluster, and within the magnitude range of  $m_r=17$ (just below the turnoff) to $m_r=24$. The properties of our simulations are reported in Table \ref{tab:1}.

The simulations span a range of concentrations\footnote{Defined as $c=\log (R_t/R_c)$, with $R_t$ the tidal and $R_c$ the core radius.} $c=1.12-2.06$ similar to typical Galactic GCs, with a range of final binary fractions from $\simeq5$\% to $\simeq30$\% (consistent with observations, e.g., \citealp{Milone2012,Sollima2007}). Note that we aim to explore a significant range of the parameter space without tuning our simulations to any specific GC.

Since the binary fraction is a radially-dependent quantity, in Table \ref{tab:1} we report three values for the 11 Gyr snapshots: the global binary fraction, the one within the FoV of 4 arcmin and with a magnitude cut between 17 and 24 $r$-band magnitudes, and the one within the core radius $R_c$. Note that, interestingly, the binary fraction within the core radius for the simulations with 50\% primordial binaries settles to the value of $\simeq30$\%, independent of the concentration of the cluster.

In a separate paper \citep{Bianchini2015d}, we investigate in detail the state of partial energy equipartition reached by our simulations and its relation to the clusters properties. Here we focus on the fact that all the simulations display mass-dependent kinematics.

\subsection{Incorporating errors in the simulation output}
\label{sec:err}
To enable a comparison with observations, we add realistic errors to the simulations. The errors for each star are selected to reproduce the trends of the errors in HSTPROMO data (e.g., Figure 18 from \citealp{Bellini2014}). For both proper motions and magnitudes, the errors are modeled to be magnitude dependent by assigning larger errors to fainter stars.\footnote{In real \textit{HST} proper-motion observations, the errors are also radial dependent, with higher errors for stars in the crowded center; this effect is not considered here.} The errors are drawn at random from a Gaussian distribution\footnote{In the observations, photometric errors are non-symmetric; this can make some main-sequence stars appear as red as binaries. In our simulations we do not consider this possibility, assuming that these apparent binaries are reliably rejected from observations \citep{Bellini2014}.} with standard deviation increasing with stellar magnitude, following a growth $a/(b-\mbox{mag})$, where $a$ and $b$ are such that the errors increase asymptotically for mag=25.5 in the $i-$band, and produce an average magnitude error (average for all the stars) of $\approx0.02$ and of $\approx0.1$ mas yr$^{-1}$ for proper motions (corresponding to $\approx2$ km s$^{-1}$ at the distance\footnote{The proper motion are reported in mas yr$^{-1}$; assuming a distance $d$ in kpc, a multiplication of 4.74$d$ yields the values in km s$^{-1}$.} of 5 kpc).
Figure \ref{fig:2}, shows an example of a color-magnitude diagram with and without errors. Note that the detailed shape of the color-magnitude diagram does not reproduce the typical observed r vs. r-i diagram.

\subsection{Construction of kinematic profiles}
We construct kinematic profiles by dividing the projected data into concentric radial bins each containing an equal number of stars. For every bin, a maximum-likelihood estimator \citep{PryorMeylan1993} is used to obtain the velocity dispersion and the associated error, taking into account the individual measurement errors. The proper motions are decomposed into radial and tangential components on the plane-of-the-sky and the average dispersion $\sigma=\sqrt{(\sigma_R^2+\sigma_T^2)/2}$ is used.

\section{The effect of unresolved binaries on kinematics}
\label{sec:3}

\begin{figure*}[p]
\centering
\includegraphics[width=0.85\textwidth]{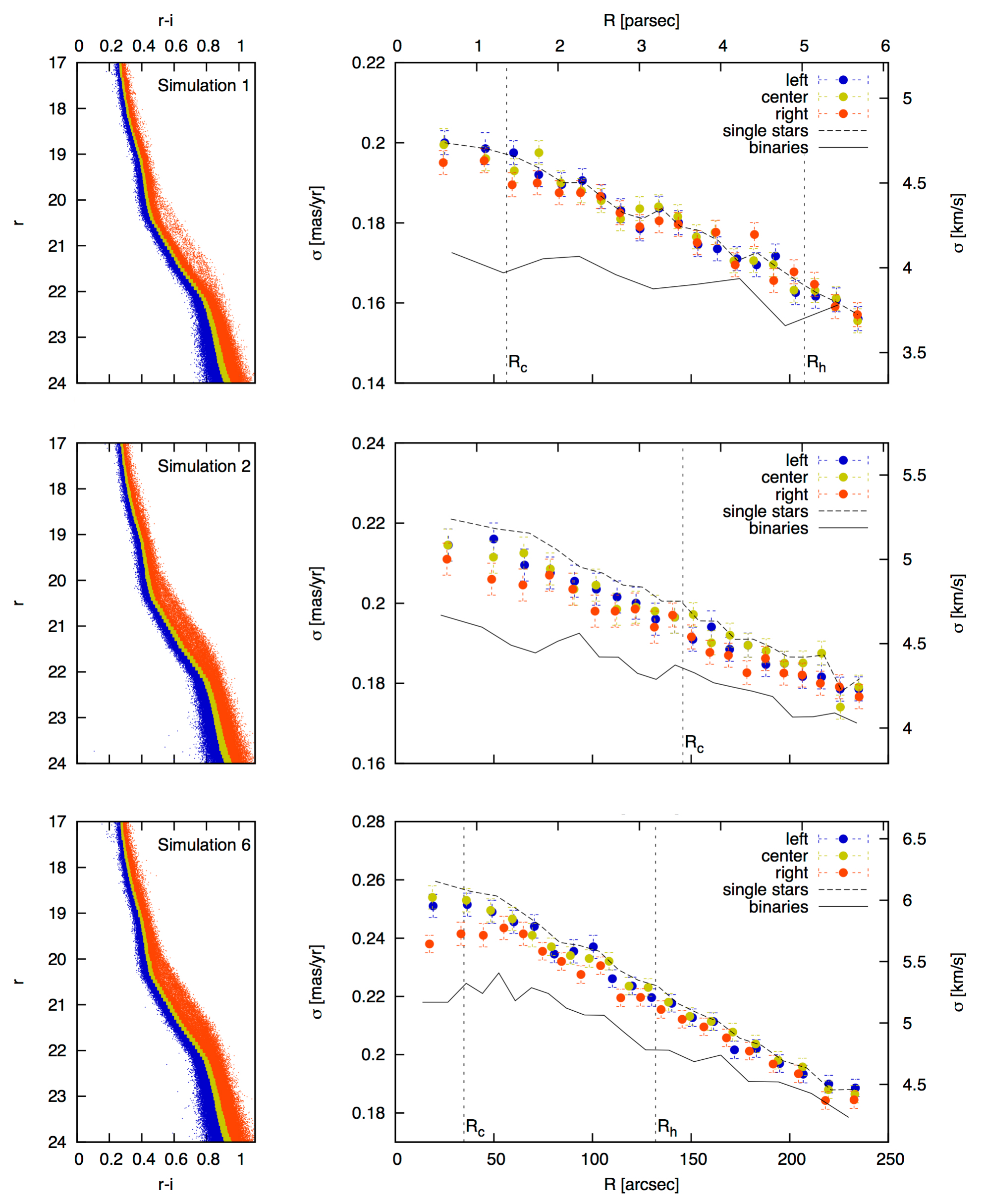}
\caption{\textbf{Left:} Color magnitude diagrams for the main sequence of our simulated data divided into three color bins, each containing 1/3 of the stars. \textbf{Right:} Velocity-dispersion profiles for the left, center and right color bins (blue, yellow, red, respectively). The profiles are constructed without distinguishing between single stars and binaries. The velocity-dispersion profiles for single stars and for binaries are shown by dashed and solid black lines, respectively. The vertical dotted lines indicate the core radii $R_c$ and the half-light radii $R_h$. Simulations 1, 2 and 6 are shown. The first shows no difference in velocity dispersion with color, while the second and the last, characterized by a high binary contamination $\simeq50$\% in the right bins (see Table \ref{tab:2}), show significant kinematic differences (up to 0.35 km s$^{-1}$, corresponding to 6\% of the central velocity dispersion) that become more pronounced in the center due to increasing binary fractions due to mass segregation.}
\label{fig:3}
\end{figure*}

\begin{table*}
\begin{center}
\caption{}
\begin{tabular}{lccccccccc}
\hline\hline
Simulation & \multicolumn{2}{c}{velocity dispersion} & difference & \multicolumn{3}{c}{binary fraction}&\multicolumn{3}{c}{velocity dispersion} \\
 & mas yr$^{-1}$ & mas yr$^{-1}$ & \% & \multicolumn{3}{c}{\%} & \multicolumn{3}{c}{mas yr$^{-1}$}\\
&single stars&binary stars&  & left & center & right  & left & center & right\\

\hline
Simulation 1		& 0.1804$\pm$0.0004 &0.166$\pm$0.002& 8.0 &1.4  &2.4 &12.5	&0.180$\pm0.001$ & 0.180$\pm0.001$ & 0.180$\pm0.001$\\
Simulation 2		& 0.1995$\pm$0.0005 &0.183$\pm$0.001& 8.2 & 12.7&18.4 &54.4	& 0.196$\pm0.001$& 0.196$\pm0.001$ &0.193$\pm0.001$\\

Simulation 3		& 0.1488$\pm$0.0005 &0.137$\pm$0.001& 7.9 & 4.3&5.0 &19.6	& 0.148$\pm0.001$&0.148$\pm0.001$ &0.148$\pm0.001$\\
Simulation 4		& 0.1670$\pm$0.0006 &0.158$\pm$0.001& 5.4 & 16.1&22.7 &59.0	& 0.165$\pm0.001$&0.164$\pm0.001$ &0.164$\pm0.001$\\

Simulation 5		& 0.2110$\pm$0.0004 &0.199$\pm$0.002& 5.7 & 2.0& 3.0&15.5	& 0.211$\pm0.001$& 0.210$\pm0.001$&0.210$\pm0.001$\\
Simulation 6		& 0.2255$\pm$0.0004 &0.205$\pm$0.001& 9.1 & 6.4& 11.3&45.7	&0.223$\pm0.001$  & 0.224$\pm0.001$ & 0.220$\pm0.001$\\
\hline
\end{tabular}
\tablecomments{From column 1-3: Velocity dispersions for single stars and binaries within the FoV and the corresponding percentage difference. Binaries have a lower velocity dispersion due to the effect of energy equipartition. From column 4-9: Binary fraction and velocity dispersion as a function of color bin for the simulations with errors included. The global values of velocity dispersion are consistent for different color bins, except for the more concentrated simulations with high binary fractions (Simulation 2 and 6).}
\label{tab:2}
\end{center}
\end{table*}

\begin{figure*}[t]
\centering
\includegraphics[width=0.95\textwidth]{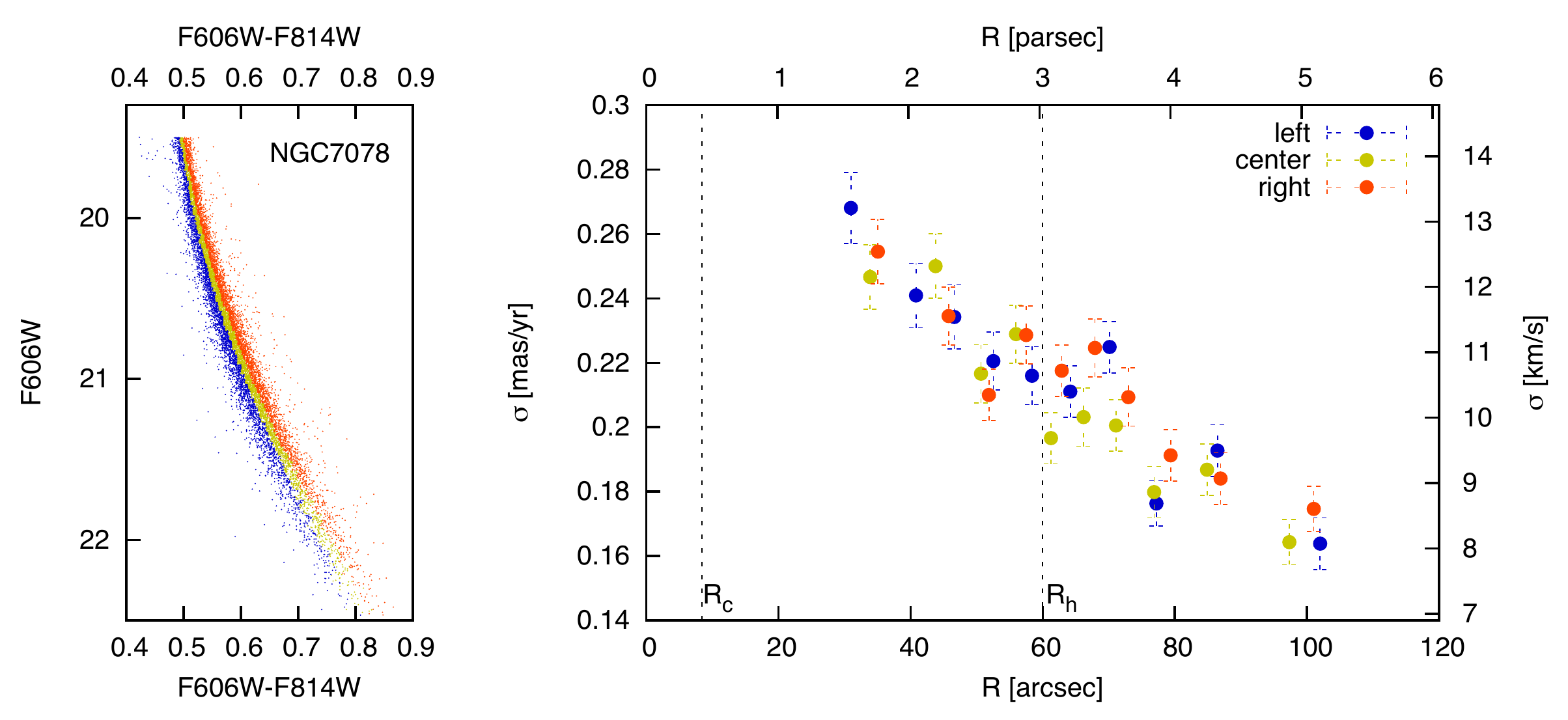}

\caption{\textbf{Left:} Color magnitude diagram for the main-sequence stars of NGC 7078 with high-quality measured HSTPROMO proper motions from \citet{Bellini2014} (photometry from \citealp{Anderson2008}). Each color bin contains $\approx4000$ stars. \textbf{Right:} Velocity-dispersion profiles for each of the color bins as in Fig. \ref{fig:3}. The three profiles are consistent with each other and do not show any kinematic signature ascribable to the presence of binary stars, consistent with the low binary fraction measured for NGC 7078 of $3\pm1$\% in the observed field-of-view (\citealp{Milone2012}).}
\label{fig:4}
\end{figure*}

We analyze the differences in kinematics between binaries and single stars, using the proper-motion samples extracted from our simulations, restricted within the FoV of 4 arcmin with added velocity errors.

First, we assume that we are able to distinguish single stars from binaries and construct separately their velocity-dispersion profiles. In the right panel of Figure \ref{fig:2}, we show the result for Simulation 1, where binaries show a lower velocity-dispersion profile than single stars. For completeness, we show the velocity dispersion calculated for the magnitude range typically measured in line-of-sight velocity samples (bright stars), and for typical proper-motion samples. Note that, for line-of-sight measurements, the main bias introduced by binaries is an overestimate of the velocity due to the motions of the stellar components around their mutual barycenter (e.g., \citealp{Minor2010}). Here we focus only on the effect on proper-motion samples, which so far has remained unexplored.

For each simulation we calculate the global value of velocity dispersion for binaries and single stars within the FoV and their percentage difference. The differences range between $\simeq5\%$ and $\simeq9\%$. The values are reported in Table \ref{tab:2}.

We now consider the more realistic case in which we cannot distinguish binaries from single stars. Since unresolved binaries are seen as single stars with fluxes given by the sum of two stellar components, we expect them to be located preferentially on the redder side of the main sequence (see Figure \ref{fig:2}). We divide the main sequence into three color bins restricted to the FoV of 4 arcmin and with $r$-band magnitudes between 17 and 24 mag. We label them as left, center, and right bins, each containing 1/3 of the stars ($\approx20,000$-50,000 stars each, depending on the simulation). We calculate the fraction of binaries contaminating each color bin and the corresponding velocity dispersion without distinguishing binaries from single stars. We report the values in Table \ref{tab:2}.

As expected, the right (redder) bins have a higher contamination of binaries, up to $\simeq60$\%. However, the velocity dispersions for the different color bins are consistent with each other, except for Simulation 2 and Simulation 6. These two simulations are, in fact, characterized by high binary contamination in the right color bin (54.4\% and 45.7\%, respectively) and show a small ($\simeq0.1$ km s$^{-1}$, 1.5\% of the velocity dispersion) but significant difference in the global velocity dispersion at the $2\sigma$ level. In contrast, Simulation 4, even though characterized by a 59.0\% binary contamination in the right bin, does not display color-dependent kinematic differences. This is explained by the fact that this simulation is the least concentrated ($c=1.12$) and reaches a low level of partial energy equipartition, and hence displays a weak kinematic-dependence on stellar mass (see \citealp{Bianchini2015d}).

We note that the differences obtained in the velocity dispersion depend on the assumption for the photometric errors: assuming a more accurate photometry would enable us to distinguish binaries from single stars more efficiently. This would allow us to measure larger differences in velocity dispersion between the color bins.

Figure \ref{fig:3} shows a comparison between the velocity-dispersion profiles of the three color bins, and the corresponding profiles for single stars and binaries only. For Simulations 2 and 6, the kinematic difference becomes larger in the central region, where the binary fraction increases. In particular, in the central region of Simulation 6, the right color bin differs from the left and the center bins by $\approx0.015$ mas yr$^{-1}$ (0.35 km s$^{-1}$ at 5 kpc, corresponding to $\approx6$\% of the central velocity dispersion), with a significance of $3.5\sigma$.

\section{Application to \textit{HST} proper motions of NGC 7078}
\label{sec:4}

We apply the analysis described above to the HSTPROMO proper motions of NGC 7078 \citep[M15,][]{Bellini2014}, consisting of 74,831 stars in a field of $\approx180\times195$ arcsec$^2$ around the center of the cluster, for which color information from the F606W and F814W bands are available. We choose NGC 7078, since a detailed understanding of the sample's quality is already available and discussed in \citet[]{Bellini2014}.

We perform the analysis on a high-quality subsample of the data in which the contaminant stars and mismatched proper motions are rejected, and only highly-accurate proper motions are considered. We follow the selection criteria outlined in Section 7.5 of \cite[]{Bellini2014} and restrict our analysis to the main-sequence stars between 19.5-22.5 F606W magnitude. The final sample consists of 12,027 stars with an average proper-motion error of 0.07 mas yr$^{-1}$ (corresponding to 3.4 km s$^{-1}$ at 10.4 kpc, \citealp[2010 edition]{Harris1996}).

Figure \ref{fig:4} shows the velocity-dispersion profiles for the color bins. The global velocity dispersions are $0.214\pm0.003$, $0.207\pm0.003$, $0.213\pm0.003$ mas yr$^{-1}$, for the left, central, and right bins respectively, measured with an accuracy of $\sim1.5\%$. The maximum difference is found between the central and left bin, corresponding to 2\% of the velocity dispersion, and the velocity dispersions are still consistent within less than $2\sigma$. Therefore, we do not detect any significant difference between the profiles, ascribeble to the presence of binaries. This is consistent with the low binary fraction of $3\pm1$\% measured in NGC 7078 (\citealp{Milone2012}) similar to our Simulation 1, where binaries do not lead to any kinematic signatures.

Finally, we note that NGC 7078 is known to host at least three stellar populations (e.g. \citealp{Piotto2015}) differing in helium abundances (and therefore main-sequence colors). The lack of velocity dispersion differences between the color bins (Fig. \ref{fig:4}) can be interpreted in the context of multiple stellar populations as an indication of the absence of significant kinematic differences among them, in the inner $\simeq2$ arcmin.
\section{Conclusions}
\label{sec:5}
We investigated and quantified the effect of unresolved binaries on GC proper-motion dispersion profiles using Monte Carlo cluster simulations. The simulations cover a large range of concentrations and binary fractions (from $\simeq5$\% to $\simeq30$\%) and reach a state of partial energy equipartition that imprints a lower velocity dispersion on binary stars (that are, on average, more massive kinematic tracers than the single stars). From the analysis of the simulations, treated similarly to \textit{HST} proper-motion samples, we conclude the following:

\begin{itemize}
\item Binaries can introduce a color-dependent bias in the velocity dispersion calculated for the main sequence, with the reddest stars showing a lower velocity dispersion, due to the higher fraction of contaminating binaries. Only simulations with a high binary fraction (initial binary fraction of 50\%) and a high concentration (i.e., more efficient in reaching a state closer to energy equipartition, hence displaying a stronger mass-dependence of their kinematics, \citealp{Bianchini2015d}) show a significant difference in velocity dispersion, of the order of 0.1-0.3 km s$^{-1}$ (Simulation 2 and Simulation 6), corresponding to 1-6\% of the velocity dispersion. The effect is larger in the center where the binary fraction increases. The low level of the bias indicates that proper-motion data are less affected by binaries than typical line-of-sight samples. Note that the color-dependent bias due to the presence of binaries is a generic result, merely due to the presence of a stellar component more-massive than average on the red side of the main sequence; however, the quantitative details could depend on the specific initial conditions adopted, in particular on the ability of efficiently distinguish binaries from single stars, the shape of the color-magnitude diagram, and the photometric errors.

\item With state-of-the-art \textit{HST} proper-motion data (\citealp[]{Bellini2014,Watkins2015,Watkins2015b}) it is possible to measure such low kinematic differences since sufficient stars are available to achieve low random errors. However, at this very low error level, other systematic effects can influence the measurements (e.g., contaminant stars, mismatched proper motions, or astrometric blends, \citealp[]{Bellini2014}), making the detection of the kinematic effects of binaries challenging. This suggests that the kinematic impact of binaries, quantified in our work, must be taken into account in the error budget of any proper-motion analysis of \textit{HST} data. This is particularly important in the context of multiple stellar populations, where different color bins could contain different stellar populations, with possible intrinsically different kinematics (see \citealp[]{Henault-Brunet2015,Bellini2015,Richer2013}).

\item We applied our analysis to the high-quality HSTPROMO dataset of NGC 7078. The velocity dispersions are measured with an accuracy $\sim1.5\%$ and we confirm that no kinematic effects due to unresolved binaries is detectable, consistent with the predictions from our simulations for a low binary fraction GC ($<5$\%). This analysis can be interpreted in the context of multiple stellar populations as an indication of the lack of significant kinematic differences among the stellar populations of NGC 7078.
\end{itemize}

\acknowledgments
We are grateful to Jonathan M. B. Downing for providing the Monte Carlo simulations used in this work. PB thanks Anna Sippel for interesting comments and discussions. AB acknowledges support from HST grant AR-12845, provided by the Space Telescope Science Institute, which is operated by AURA, Inc., under NASA contract NAS 5-26555. We thank the referee for helping improving the quality of our manuscript.

\end{document}